\documentclass[amsmath,preprint,amssymb,onecolumn,pre,longbibliography]{revtex4-1}

\usepackage{graphicx}
\usepackage{dcolumn}
\usepackage{bm}
\usepackage{color}

\begin{document}

\title{Scaling laws of top jet drop size and speed from bubble bursting including gravity and inviscid limit}

\author{Alfonso M. Ga{\~n\'a}n-Calvo}
\email{amgc@us.es}
\affiliation{Dept. Ingenier{\'\i}a Aerospacial y Mec{\'a}nica de Fluidos,
Universidad de Sevilla.\\
Camino de los Descubrimientos s/n 41092, Spain.}

\date{\today}

\begin{abstract}
Jet droplets from bubble bursting are determined by a limited parametrical space: the liquid properties (surface tension, viscosity, and density), mother bubble size and acceleration of gravity. Thus, the two resulting parameters from dimensional analysis (usually, the Ohnesorge and Bond numbers, Oh and Bo) completely define this phenomenon when both the trapped gas in the bubble and the environment gas have negligible density. A detailed physical description of the ejection process to model both the ejected droplet radius and its initial launch speed is provided, leading to a scaling law including both Oh and Bo. Two critical values of Oh determine two limiting situations: one (Oh$_1$=0.038) is the critical value for which the ejected droplet size is minimum and the ejection speed maximum, and the other (Oh$_2$=0.0045) is a new critical value which signals when viscous effects vanish. Gravity effects (Bo) are consistently introduced from energy conservation principles. The proposed scaling laws produce a remarkable collapse of published experimental measurements collected for both the ejected droplet radius and ejection speed.
\end{abstract}


\maketitle

Bubble bursting is a particular case of a general class of free surface axisymmetric capillary flows producing unsteady liquid ejections. Yarin \cite{Yarin2006} discussed several related phenomena (droplet impact, film breakage in bubble bursting, etc.) where a sudden change in the overall potential energy of the system leads to the radial progression and collapse of a wave package \cite{Worthington1897,Zeff2000}.  Those phenomena plague the dynamics of free surface flows at length scales comparable to capillary lengths. Bubble bursting at the liquid surface may arise as a consequence of trapped or dissolved gas reaching the surface, but also bubble trapping caused by the axisymmetric wave collapse after a droplet impact on a liquid surface produces subsequent microdroplet ejection after the initial large scale jetting. At planetary scales, the largest free surface between liquid and gas is the sea surface, where the dynamical interaction between these phases involve scales spanning about ten orders of magnitude (from tens of nanometers to hundreds of meters). Yet, the mixing and penetration of each phase in the other (in the form of droplets or bubbles) is dominated by the capillary lengths and below. Indeed, at the smaller scales of capillary phenomena, very small droplets are always released: this peculiar feature is so fundamental that it largely determines the global dynamics of the gas phase (atmosphere) through the continuous formation of large masses of aerosols from ocean spray\cite{Veron2015}. These aerosols form the cloud condensation nuclei (CCN) that eventually regulate precipitations and the radiant balance of the earth.

Among the different spray formation mechanisms, what is known as {\sl bubble jetting} was early identified as the one producing the smaller droplets that reach farther away from the free surface, due to the vigorous ejection taking place perpendicularly from that surface. That ejection was early observed and reported in detail by Worthington \cite{Worthington1897}, and subsequently attracted much attention from climate scientists \cite{Kientzler1954,Blanchard1957,Hayami1958,MacIntyre1972,Blanchard1989,Spiel1995,Spiel1997}. Bubble jetting entails the collapse of a capillary wave package onto the axis of symmetry and the eventual ejection of liquid along the axis of symmetry, due to conservation of mass and momentum. The allure of this peculiar phenomenon comes not only from its own physical beauty, symmetry and richness beget by just a few parameters, but from its transversal impact and direct role in the global complexity and life in the planet. In fact, one can easily understand the importance of the aerosols generated in large scale phenomena like planet albedo, precipitations, or airborne microbial dissemination.

\begin{figure}[htb]
\centering
\includegraphics[width=0.65\textwidth]{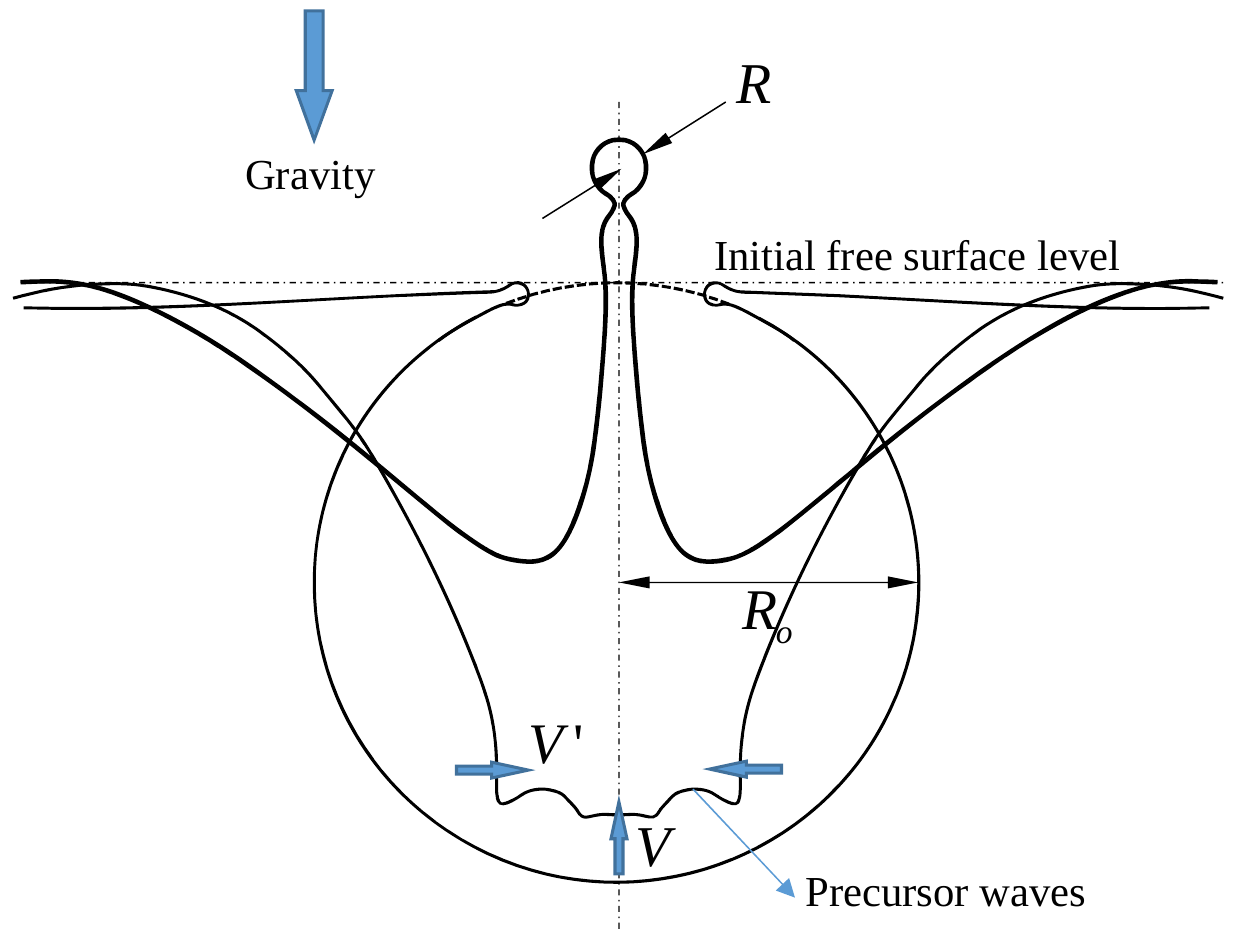}
\vspace{-3mm}
\caption{Schematics of the bubble bursting phenomenon showing three stages where the initial free surface level, initial bubble radius $R_o$, ejected droplet radius $R$, main wave speed $V'$, and induced ejection speed $V$ are indicated. The profiles shown correspond to three illustrating but arbitrary times selected from the case Oh$=0.01$ and Bo$=0.01$ in a detailed numerical simulation \cite{Deike2018}.}
\label{fig1}
\vspace{-3mm}
\end{figure}

This work analyzes in detail bubble bursting on a surface, with the aim to provide a complete description and predicting models for the two main mechanical parameters to determine the fate of the ejecta as airborne aerosols: the size (radius) $R$ of the first ejected droplet and its initial speed $V$ (figure \ref{fig1}). In this phenomenon, the source of energy mainly comes from the breakage of a liquid film exposed to air. In a lesser extent, the sudden local imbalance of the gravity potential associated to the open cavity created right after the film breakage may also contribute to the ejection. Besides, several simultaneous droplets are in most cases formed from the breakup of the issued jet. However, since the first droplet is the one taking the most important fraction of energy from the short living jet, this work is focused on that droplet. Indeed, it is the one with larger ejection speed and highest reach.

The physics involved has been discussed by several authors who have provided successive insightful approaches \cite{MacIntyre1972,Boulton1993,Duchemin2002,Walls2015,Krishnan2017,Deike2018}. A synthesis of the existing arguments was briefly discussed in \cite{Ganan2017}: those arguments pointed to the existence of an overall speed of the capillary wave front that should be of the order of $V_o= \left(\frac{\sigma}{\rho R_o}\right)^{1/2}$. This assumes that (i) the dominant wave number $k$ should be comparable to $R_o^{-1}$, and (ii) that the wave undergoes a viscous damping rate as $t_D^{-1}\sim \mu/\left(\rho R_o^2\right)$ which should be smaller than the inverse of the time of collapse of the wave $t_o^{-1}\sim \left(\frac{\sigma}{\rho R_o^3}\right)^{1/2}$. In other words, one should have $t_D>t_o$, which immediately implies that the  Ohnesorge number $\text{Oh}=\frac{\mu}{(\rho\sigma R_o)^{1/2}}$ should be below a critical one (here, $\text{Oh}_1$) to have a {\sl sufficiently energetic} jetting for droplet ejection. That critical number $\text{Oh}_1$ was experimentally calculated by Walls et al. \cite{Walls2015}, including the influence of the gravity using the Bond number $\text{Bo}=\rho g R_o^2/\sigma$. In the limit of very small Bo numbers, they obtained $\text{Oh}_1\simeq 0.037$, which was confirmed in \cite{Krishnan2017}.

However, identifying the critical $\text{Oh}_1$ does not resolve the dependency that both the ejected droplet size and its speed should have on the three relevant physical properties of the liquid $\{\rho, \sigma,\mu\}$, the bubble radius $R_o$, and the acceleration of gravity $g$. Following dimensional analysis, that dependency should be given in terms of two non-dimensional variables, for example in the forms $R/l_\mu=f_R\left(\text{Oh},\text{Bo}\right)$ and $V/V_\mu=f_V\left(\text{Oh},\text{Bo}\right)$, where $l_\mu=\mu^2/(\rho\sigma)$ and $V_\mu=\sigma/\mu$ are the capillary-viscous length and speed, respectively. The interested reader can find the whole formulation of this work in alternative terms of $R/R_o=f_R\left(\text{Oh},\text{Bo}\right)$ and $V/V_o=f_V\left(\text{Oh},\text{Bo}\right)$ in the Supplementary Material \cite{SM}. In the following and for illustrative purposes, the obtaining and limits of applicability of the scaling proposed in \cite{Ganan2017} is outlined from physical principles. This work aims to provide a valid scaling for $R$ and $V$ in the whole parametrical space $\{$Oh, Bo$\}$ where experiments, numerical simulations and limiting behaviors have been reported in the literature. In particular, in the revision of this work, another paper \cite{Brasz2018} has discussed the minimum size of the drops ejected, which is also contemplated here for completeness.

Among others, Krishnan et al. \cite{Krishnan2017} neatly described (see their figure 8) how the different wavelengths $\lambda_i$ of the wave packet produced by the breakup of the liquid film sequentially arrive at the axis segregated by their different wave speeds $\left(\sigma/(\rho \lambda_i)\right)^{1/2}$. In \cite{Ganan2017}, I observed that when the front of the main capillary wave producing ejection collapses at the axis, the curvature reversal of the surface involved in the onset of ejection imply that all terms of the momentum equation should be locally comparable.
In brief, the collapse of a wave with speed $V_L$ and amplitude $L$ leading to the ejection of a mass with characteristic radial size $R$ and axial speed $V$ should obey the dimensional balance:
\begin{equation}
O\left(\rho V^2/L\right) \sim O\left(\mu V_L L^{-2}\right) \sim O\left(\sigma R^{-2}\right)
\label{momentum}
\end{equation}
which together with the conservation of mass, i.e.
\begin{equation}
O\left(V R^2\right)\sim O\left(V_L LR\right),
\label{mass}
\end{equation}
leads to:
\begin{eqnarray}
R/l_\mu \sim \left(V/V_\mu\right)^{-5/3},\\
L/l_\mu\sim \left(V/V_\mu\right)^{-4/3},\\
V_L/V_\mu \sim \left(V/V_\mu\right)^{2/3}.
\label{RLV}
\end{eqnarray}
In reality, the scaling relationships (\ref{RLV}) hold for every wave with arbitrary wavelength $L$ that successfully arrives at the axis, be that wave one of the precursor wavelets or the main wave with a wavelength comparable to $R_o$. Indeed, when Oh is sufficiently small, the precursor waves segregate according to their wave speed $V_L$, forming the capillary ripplets studied by different authors\cite{Crapper1957,Longuet-Higgins1963, Longuet-Higgins1992}: each individual wavelet arrival from the precursor wave pack \cite{MacIntyre1972,Boulton1993,Krishnan2017,Deike2018} produces its own collapse with curvature reversal and  partial ejection, (and often a tiny bubble entrapment) which is overcome by the more energetic wave leading to first successful ejection. For example, Deike et al \cite{Deike2018} neatly show in their figure 4b the appearance of more than one subsequent velocity peaks at the axis. In this sense, the wave collapse sequence observed is akin to a race among small, fast but weak devices and larger, slower but stronger ones: at some point, one of them has the right balance of velocity and strength to prevail. In the vast majority of cases, the slower but stronger waves produce the droplet ejection. A salient feature observed at the collapse of the capillary wave pack at the axis is that the amplitude of the waves appear comparable to their wavelength (see \cite{MacIntyre1972}, and \cite{Deike2018} figure 4a). Close to collapse, a wave is akin to a hydraulic jump or shoulder that often engulfs a small bubble after collapse.

For any wavelength $L$, one has $V_L\sim \left(\sigma /(\rho L)\right)^{1/2}$, or $\rho V_L^2\sim \sigma/L$. Moreover, given the near-zero stress condition at the surface, the strong wave leading to ejection  would also induce a radial motion in the underlying layer of liquid with speed as $V$, such that $\rho V^2\sim \mu V_L/L$. $V$ is akin to the mass-transport velocity in the analysis of Longuet-Higgins \cite{Longuet-Higgins1992}. One fundamental remark here is that the induced velocity (or {\sl mass-transport velocity}, see \cite{Longuet-Higgins1992}, figure 11) never overcomes the wave speed, i.e.  $V$ is smaller than, or at most of the same order as $V_L$, which ensures that the mass balance $O\left(V R^2\right)\sim O\left(V_L LR\right)$ previously used is fulfilled. Indeed, one should expect that the ratio $V/V_L$ vanishes for vanishing Oh numbers: in this case, only the waves with wavelength comparable to $R_o$ which set in motion most of the liquid surrounding the bubble would produce sufficient push to eject a jet. Finally, due to conservation of momentum after collapse, the induced velocity should eventually be comparable to the axial speed, both scaling as $V$.

The reader can readily verify that the above arguments based on the wave pack collapse, i.e. $\rho V_L^2\sim \sigma/L$ and $\rho V^2\sim \mu V_L/L$ are exactly equivalent to say that all terms of the momentum equation at the location of collapse should balance, expressed as in (\ref{momentum}). Finally, the condition of an {\sl efficient collapse} entails that the wave front should induce an axial motion {\sl sufficient} to launch a liquid column at vertical distances comparable to $R_o$. This can be summarized in a global energy budget as (see \cite{Ganan2017}):
\begin{equation}
\left(\text{Oh}_1\sigma R_o^2 -\mu \left(\sigma R_o^3/\rho\right)^{1/2}\right)=k \rho V^2 R^2 R_o
\label{energy0}
\end{equation}
i.e. that the total available energy in the form of surface energy, proportional to $\sigma R_o^2$, minus the total viscous dissipation of the complete wave pack $\mu \left(\sigma R_o^3/\rho\right)^{1/2}$ should be proportional to the mechanical energy of the liquid ejected column $\rho V^2 R^2 R_o$. Observe that the potential energy of gravity was not accounted for in the balance (\ref{energy0}) initially formulated in \cite{Ganan2017}. One should also be careful at considering what is understood as the ejection speed $V$ since it varies strongly with space and time time. Most authors take velocity measurements when the jet front reaches the level of the original free surface, which supports using $R_o$ as the characteristic length for the liquid column in the right hand side of (\ref{energy0}). Besides, the constant $\text{Oh}_1$ is precisely that critical Ohnesorge number above which the viscous dissipation would overcome the available surface energy, as one may readily observe dividing the whole equation (\ref{energy0}) by $\sigma R_o^2$. Combining equations (\ref{momentum}), (\ref{mass}), and (\ref{energy0}), one obtains \cite{Ganan2017}:
\begin{equation}
\frac{R}{l_\mu}= k_d \varphi^{5/4},\, \frac{V}{V_\mu}= k_v \varphi^{-3/4},
\label{original}
\end{equation}
where $\varphi=\text{Oh}^{-2}\left(\text{Oh}_1-\text{Oh}\right)$.
$k_d$ and $k_v$ would be expected to be universal constants  under the {\sl same definite criteria} to measure $R$ and $V$, or at least have a weak dependency on Bo and Oh. From these results, one has
\begin{equation}
\frac{V}{V_L}\sim \text{Oh}^{1/2}\left(\text{Oh}_1-\text{Oh}\right)^{-1/4}
\end{equation}
As anticipated, $V/V_L$ vanishes for vanishing Oh, providing consistent support to all prior assumptions. This means that in the limit Oh$\rightarrow 0$, one should expect a significant deviation from the scaling proposed in \cite{Ganan2017}, since in this limit the large wavelength waves would always take over as experimentally observed.

In summary, we have two possible causes of deviation: (i) very small Oh values, and (ii) non small Bo values. This work is dedicated to unveil the parametric dependency of these deviations from prior scaling.

About 350 published experimental and numerical data since 1954 (see table \ref{tab5}) have been analyzed\cite{Kientzler1954,Blanchard1957,Hayami1958,MacIntyre1972,Blanchard1989, Spiel1995, Spiel1997,Duchemin2002,Walls2015,Ghabache2016,Ghabache2016a}. The liquid properties are listed in table \ref{tab5}. A first important remark here is that we are considering the scaling laws for the ejection of the {\sl first} drop (or {\sl top jet drop}), which entails univalued universal constants. This does not exclude the ejection of other subsequent differently sized droplets; in particular, one can observe how the first smallest waves eject a first small drop if they are sufficiently energetic, while the last wave may also eject a large drop (see \cite{Krishnan2017}, figures 15 and 16). Second, we use experimental data where the authors measure the velocity of ejection when the jet front reaches the free surface; we call this $V$, while the final ejection velocity of the droplet (right at pinch-off, as considered by Deike et al. \cite{Deike2018}) will be called $V_j$. Naturally, one should expect  $V_j<V$, as shown by experiments and detailed numerical simulations \cite{Duchemin2002,Deike2018}.

\begin{table}
\begin{tabular}{|c|c|c|c|c|}
\hline Liquid & Ref. & $\rho$ (kg$\cdot$m$^{-3}$) &  $\sigma$ (N m$^{-1}$) & $\mu$ (Pa s) \\ \hline
SW 20$^o$C    & \cite{Blanchard1957,Blanchard1989}    & 1025  &   0.0734 &  0.00108  \\
SW 4$^o$C & \cite{Blanchard1989}   & 1028  &   0.0755 &  0.00167  \\
SW 16$^o$C  & \cite{Hayami1958}   & 1025  &   0.0736   &  0.00112  \\
SW 30$^o$C  & \cite{Hayami1958, Spiel1997}   & 1024  &   0.071  &  0.00098  \\
Water & \cite{Kientzler1954,MacIntyre1972,Spiel1995,Ghabache2016a}   & 1000  &   0.072  &  0.001  \\
W+30\% G 25$^o$C & \cite{Ghabache2016a}   & 1078  &   0.067 &  0.0021  \\
W+50\% G 25$^o$C & \cite{Ghabache2016a}   & 1130  &   0.065 &  0.0044  \\
W+60\% G 30$^o$C & \cite{Ghabache2016a}   & 1156  &   0.064 &  0.0062  \\
W+60\% G 25$^o$C & \cite{Ghabache2016a}   & 1156  &   0.064 &  0.0074  \\
W+60\% G 20$^o$C & \cite{Ghabache2016a}   & 1156  &   0.064 &  0.0097  \\
\hline
\end{tabular}
\caption{Liquid properties from experiments in the literature since 1954. SW: seawater. W + G: water-glycerol mixtures. The properties from Ghabache et al. \cite{Ghabache2016} are provided in their Table 1.}
\label{tab5}
\end{table}

For the first drop, the scaling laws (7) and (8) in \cite{Ganan2017} showed a very good agreement with experiments for Bond numbers $\text{Bo}<0.1$. However, as anticipated, the interested reader can observe apparent deviations from the alternative form of the scaling laws given by equations (4) and (5) in Supplementary Material \cite{SM} for both very small Oh and Bo of the order unity.

\begin{figure}[htb]
\centering
\includegraphics[width=0.76\textwidth]{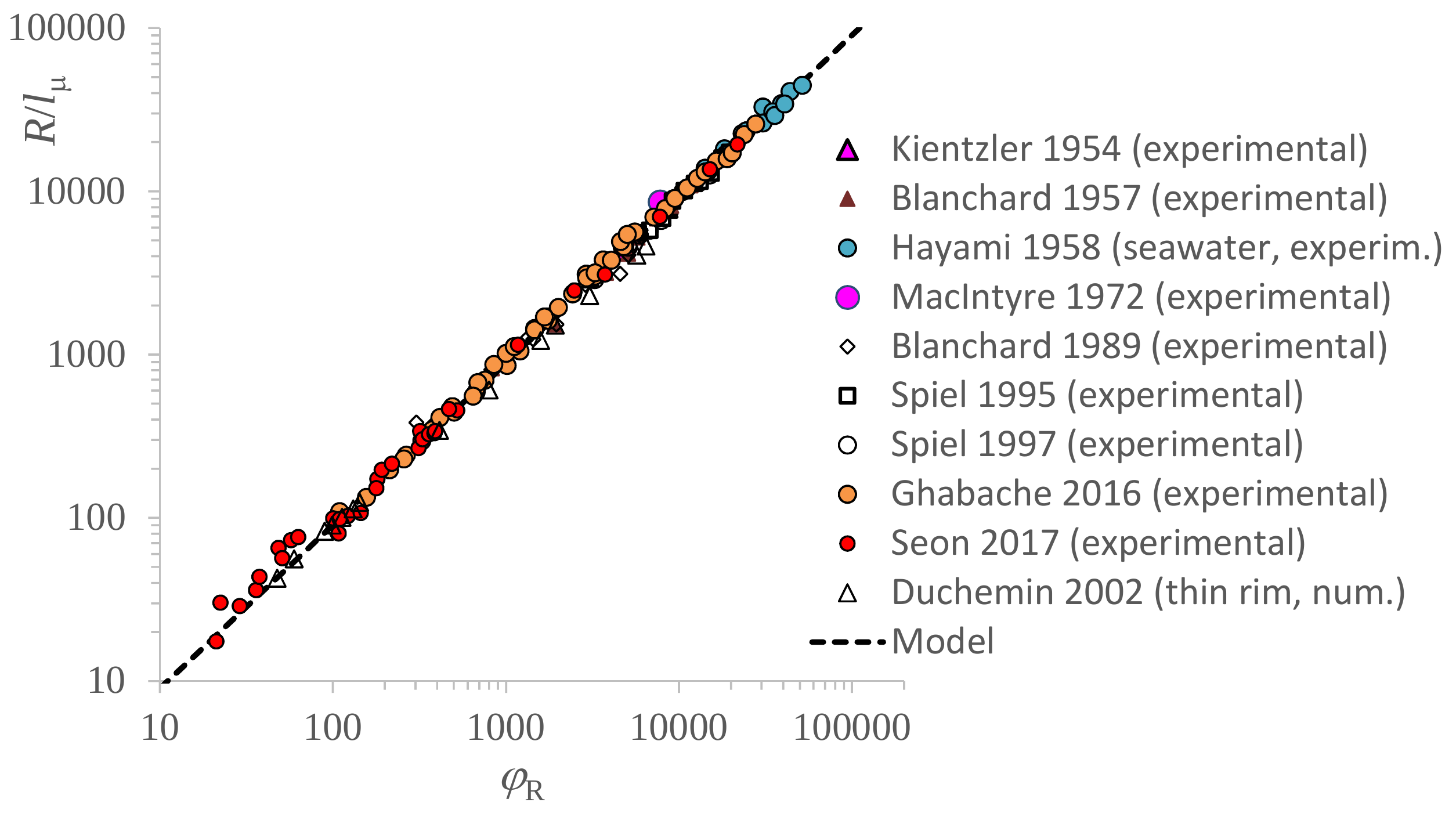}
\vspace{-3mm}
\caption{Non-dimensional droplet radius  $R/l_\mu$ as a function of the scaling variable $\varphi_{\text{\tiny R}}$, from the same data as in figure 5 in \cite{Ganan2017}. The interested reader can assess a small deviation for large values of $\varphi$ in that prior figure which is largely emphasized using the variable $R/R_o$ instead, as shown in the Supplementary Material \cite{SM}.}
\label{figR}\vspace{-3mm}
\end{figure}

Deike et al. \cite{Deike2018} made an exhaustive numerical analysis on the dynamics of the ejected jets, proposing a correction of the form $k_v\left(\text{Bo}\right)=\text{Oh}^{-3/4}_D\left(1+\alpha \text{Bo}\right)^{-3/4}$ for the scaling law (8) in \cite{Ganan2017} when $\text{Bo}>0.1$, with a critical Ohnesorge number $\text{Oh}_{D}=\text{La}^{-1/2}_*=0.045$ (Ga{\~n\'a}n-Calvo previously obtained a critical value $\text{Oh}^*=0.043$), and $\alpha=2.2$. Deike's proposal improves significantly the dispersion observed (compare figures 1 and 2 in the Supplementary Material \cite{SM}). However, that proposal does not address simultaneously the outstanding issues for both $\text{Oh}\rightarrow 0$ and non small Bo. To do so, we propose:
\begin{enumerate}
\item The induced momentum $\rho V^2$ comes from {\sl both} the faster wave by viscous mechanisms, i.e. $\mu V_L/L$, and from the final inertial push of the largest wave, i.e. $\rho V_o^2$. This can be formulated as:
    \begin{equation}
    \rho V^2 \sim \mu V_L /L + k_2 \rho V_o^2
    \label{aV}
    \end{equation}
    where the constant $\text{Oh}_2$ is called this way because it will indeed have that specific physical meaning: it will signal the small limiting value of $\text{Oh}$ below which the inertial push of the large wave takes over. It is expected to have a universal value for this problem. Retracing the same steps as before, one arrives to the following scaling expression:
\begin{equation}
\frac{V}{V_\mu}\sim \left(1+\frac{k_2}{\text{Oh}} \left(\frac{V R}{V_o R_o}\right)^3
\right)^{1/5}\left(\frac{R}{l_\mu}\right)^{-3/5},
\label{RVmu}
\end{equation}
\item The gravity potential imbalance $\rho g R_o$ created by the cavity after the film burst should be taken into account as an additional asset of energy proportional to $(\rho g R_o)R_o^3$ for the ejection. This should be formulated as an augmented version of equation (\ref{energy0}):
    \begin{equation}
    \text{Oh}_1\sigma R_o^2 -\mu \left(\sigma R_o^3/\rho\right)^{1/2}+k_{\text{\tiny Bo,1}} (\rho g R_o) R_o^3=k' \rho V^2 R^2 R_o,
    \label{energy}
    \end{equation}
    or in non-dimensional form:
     \begin{equation}
    \text{Oh}_1 -\text{Oh} +k_{\text{\tiny Bo,1}}\text{Bo}=k' \frac{\rho }{\sigma R_o}V^2 R^2,
    \label{energy1}
    \end{equation}
     where both $k_{\text{\tiny Bo,1}}$ and $k'$ are expected, again, to have universal values under the {\sl same criteria} to measure $R$ and $V$. In this regard, it is worth noting that equation (\ref{energy}) assumes a balance that should hold at each point of the ejection, which entails having {\sl different} values for $k_{\text{\tiny Bo,1}}$ and $k'$ if one considers that $V$ is the jet front speed measured at the free surface or anywhere else.
     We will come back to this issue once we get to the experimental validation.
\end{enumerate}
Given that the ejected droplet radius $R$ has an unequivocal final value, while $V$ depends on the measurement criteria, we can first focus on the scaling law of $R$. For $\text{Oh}\ll \text{Oh}_1$ (i.e. the low viscosity and asymptotically inviscid cases), the product $\left(\frac{V\, R}{V_o\,R_o}\right)^3$ is asymptotically equal to a constant that we can define as $\text{Oh}_2/k_2$. In other words, as far as one considers factors of the order $\frac{\text{Oh}_2}{\text{Oh}}$, one should neglect factors of the order of $\frac{\text{Oh}}{\text{Oh}_1}$, and viceversa, i.e. retaining factors like $\frac{\text{Oh}}{\text{Oh}_1}$ leads to neglect factors of the order $\frac{\text{Oh}_2}{\text{Oh}}$. Considering this and eliminating $V$ from (\ref{RVmu}) (or from equation (8) in Supplementary Material \cite{SM}), and (\ref{energy1}), one explicitly has for the ejected droplet radius:
\begin{equation}
\frac{R}{l_\mu}\sim \frac{\left(\text{Oh}^{-1}\left(\frac{\text{Oh}_1}{ \text{Oh}}-1+k_{\text{\tiny Bo,1}} G\right)\right)^{5/4}}{\left(1+\frac{\text{Oh}_2}{\text{Oh}}
\right)^{1/2}}\equiv \varphi_{\text{\tiny R}},
\label{Rlmu}
\end{equation}
where $G=\text{Bo}/\text{Oh}$ is the ratio of gravity over viscous forces, with $k_{\text{\tiny Bo,1}}$ 
a fitting constant. One would expect that the experiments should provide universal values of the critical numbers Oh$_1$ and Oh$_2$. To this end, one may use the same experimental data set employed in \cite{Ganan2017}, including all data for Bo$>0.1$. First, Oh$_1$ and Oh$_2$ are resolved together with the fitting parameters $k_{\text{\tiny Bo,1}}$ and $k_{\text{\tiny Bo,2}}$ by any valid optimization method (e.g. minimum least squares) using measurements of the top jet  droplet radius $R$. The optimum fitting is shown in figure \ref{figR}a with Oh$_1=0.038$ (very close to Walls' critical value 0.037), Oh$_2=0.0045$, $k_{\text{\tiny Bo,1}}=0.006$. 
The scaling prefactor  such that $R/l_\mu=k_d \varphi_{\text{\tiny R}}$ results $k_d=0.9$.

The extraordinary fitting found validates the newly proposed scaling (\ref{Rlmu}). The interested reader can see an enhanced comparison between the original and the new scalings in figures 1(a) and 3(a) in the Supplementary Material \cite{SM}. The new scaling encapsulate a rich physical spectrum summarized in the following:
\begin{enumerate}
\item The number $\text{Oh}_1$ indicates the limiting value of the Ohnesorge number for which the droplet radius nearly vanishes, and below which droplet ejection appears just marginally, originating larger droplets \cite{Duchemin2002} (in the revision of this work, a minimum attainable droplet size is proposed in \cite{Brasz2018} for Oh$\rightarrow$ Oh$_1$. In reality, as discussed in \cite{Brasz2018}, dominant viscous effects should make $R/l_\mu$ minimum but nonzero at that singular point). Walls obtained $\text{Oh}_1\simeq 0.037$ while we propose $\text{Oh}_1= 0.038$ (practically indistinguishable) when $\text{Bo}\rightarrow 0$. More precisely, to this end one should have $G\rightarrow 0$.
\item The number $\text{Oh}_2$ (small compared to $\text{Oh}_1$) is the value of the Ohnesorge number below which viscous forces become negligible compared to capillary and inertia forces. The main mechanism leading to ejection becomes the collapse of the larger and slower non-linear capillary wave which inertially pushes the liquid towards the axis. In this region ($\text{Oh}<\text{Oh}_2$) the inviscid limit studied by Boulton-Stone and Blake is beautifully recovered \cite{Boulton1993} (data from their figure 4a \& 4b are used in the Supplementary Material \cite{SM}, figure 3). In this inviscid limit, expression (\ref{Rlmu}) using the viscous scaling $l_\mu$ becomes undetermined, and one should use $R/R_o$ instead (see Supplementary Material \cite{SM}). After some easy algebra, the resulting limit is a function of the Bond number alone (Oh$_1$ and Oh$_2$ are constants):
     \begin{equation}
    \frac{R}{R_o}=k_d \text{Oh}_2^{-1/2}\left(\text{Oh}_1+k_{\text{\tiny Bo,1}} \text{Bo}\right)^{5/4}.
    \end{equation}
\item In the intermediate asymptotic region $\text{Oh}_1\gg \text{Oh}\gg \text{Oh}_2$, one has
\begin{equation}
R/l_\mu\simeq k_d \text{Oh}^{-5/2}\left(\text{Oh}_1+k_{\text{\tiny Bo,1}}\text{Bo}\right)^{5/4},
\label{Rlmus}
\end{equation}
The limit described in \cite{Ganan2017} is recovered when $k_{\text{\tiny Bo,1}}G\equiv k_{\text{\tiny Bo,1}}\text{Bo/Oh} \ll 1$. Noteworthy, the effect of gravity is not due to the ascending jet, but to the gravity potential imbalance produced by the local presence of the original bubble at the surface: this is easy to understand given the much larger volume of the cavity than that of the jet, having both comparable heights.
\end{enumerate}
Now, one can use the same data as in figure \ref{figR} for the jet velocity, obtained with the same measurement criterium (whenever available). While one should expect that the values of Oh$_1$ and Oh$_2$ remain constant, one should also expect some opposing push of the gravity on the column as it rises. Therefore, one may expect variations in the values of $k_{\text{\tiny Bo,1}}$ (that can be called $k'_{\text{\tiny Bo,1}}$) and an additional term in equation (\ref{aV}) corresponding to the weight of the column for any given length that it reaches. If the criterion to measure the jet speed is when its front reaches the initial free surface of the bubble, that weight can be formulated as:
    \begin{equation}
    \rho V^2 \sim \mu V_L/L + \text{Oh}_2 \rho V_o^2-k_{\text{\tiny Bo,2}}\rho g R_o,
    \label{aV2}
    \end{equation}
Using now (\ref{aV2}), one reaches to:
\begin{equation}
\frac{V}{V_\mu}\sim\frac{\left(\text{Oh}+\text{Oh}_2- k_{\text{\tiny Bo,2}}\text{Bo}\right)^{1/2}}{\text{Oh}^{-1}\left(\text{Oh}_1-\text{Oh}+k'_{\text{\tiny Bo,1}} \text{Bo}\right)^{3/4}}\equiv \varphi_{\text{\tiny V}}.
\label{Vlmu}
\end{equation}
Thus, doing the same optimum collapse process as for the droplet radius, one effectively obtains Oh$_1=0.038$ and Oh$_2=0.0045$ (consistently with expectations from the scaling of the droplet size), and $k'_{\text{\tiny Bo,1}}=0.14$ and $k_{\text{\tiny Bo,2}}\simeq 0.004$, with a scaling prefactor $k_v=13.5$ such that $V/V_\mu=k_v \varphi_{\text{\tiny V}}$.
\begin{figure}[htb]
\centering
\includegraphics[width=0.76\textwidth]{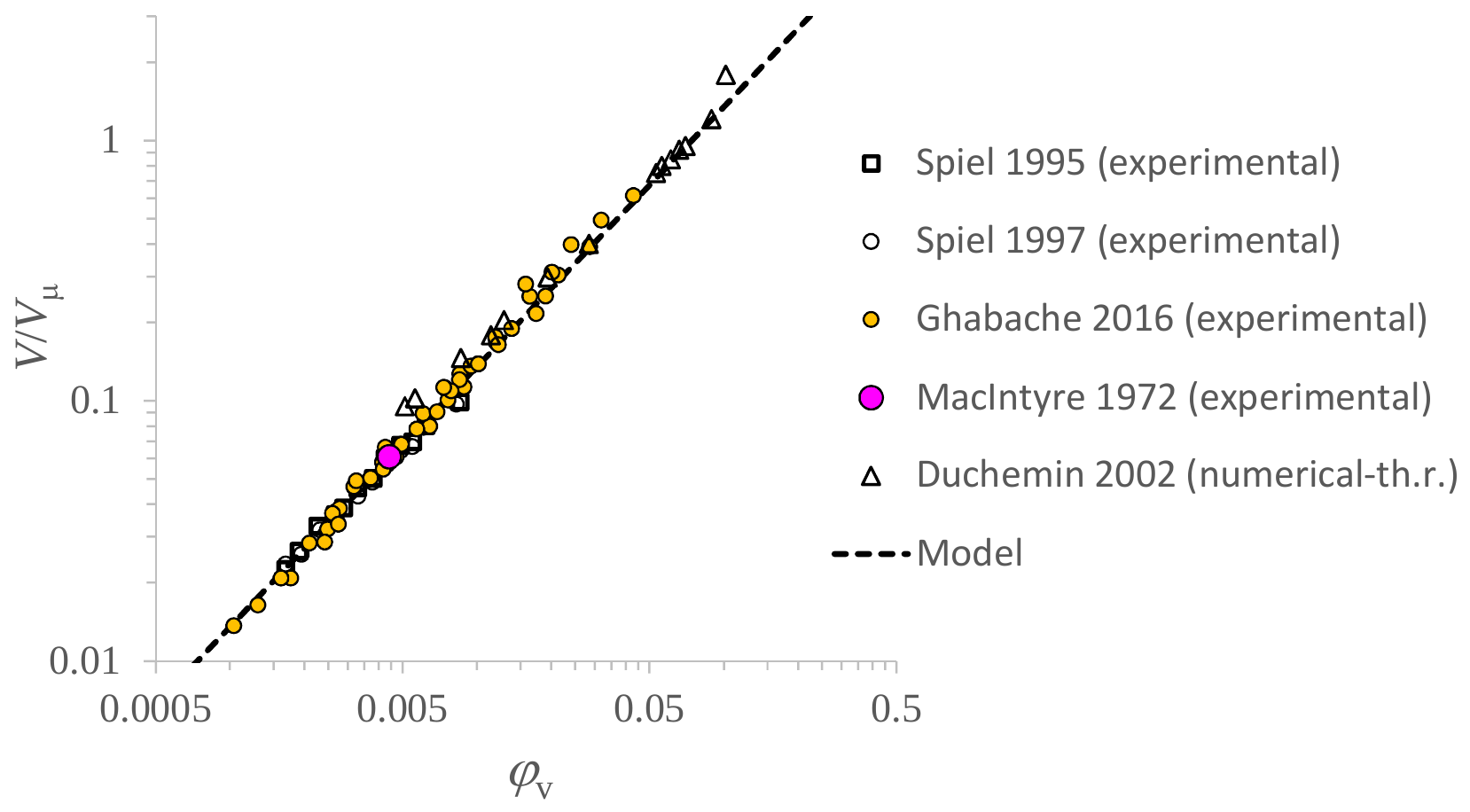}
\vspace{-3mm}
\caption{Ejection velocity at the level of the original free surface $V$ made dimensionless with $V_\mu$ (b), as a function of  $\varphi_{\text{\tiny V}}$. All available data, including those with Bo$>0.1$ (that were not included in previous study \cite{Ganan2017}), are represented.}
\label{figV}
\end{figure}
Again, a very good collapse is obtained. The interested reader can compare the data dispersion in either figures 1b or 2 with that in figure 3(b) in the Supplementary Material \cite{SM}. Observe that the inviscid limit \cite{Boulton1993} is also recovered for the jet speed, naturally (see Supplementary Material \cite{SM}). Finally, for completeness, one should also consider those ejections contemplated in \cite{Brasz2018} and previously in \cite{Duchemin2002} for values of the Ohnesorge number above the one that makes $\varphi_R$ zero. However, given the much lower ejection speeds of droplets for Oh numbers larger than Oh$_1$, for which the droplet size is minimum, the overall importance of that regime can be marginal except for the determination of the minimum ejectable droplet size \cite{Brasz2018}.

This work was supported by the Ministerio de Econom{\'\i}a y Competitividad, Plan Estatal 2013-2016 Retos, project DPI2016-78887-C3-1-R. 




%
\end{document}